\documentclass[
prl
 ,secnumarabic%
,amssymb,nobibnotes, aps, prl]{revtex4}
\usepackage{amsmath}
\usepackage{graphicx}
\input{epsf}

\newcommand{\be}{\begin{equation}}
\newcommand{\ee}{\end{equation}}
\newcommand{\bea}{\begin{eqnarray}}
\newcommand{\eea}{\end{eqnarray}}
\newcommand{\ba}{\begin{array}}
\newcommand{\ea}{\end{array}}

\renewcommand{\(}{\left(}
\renewcommand{\)}{\right)}
\def \lam{\lambda}
\def \eps{\epsilon}
\def \veps{\varepsilon}
\def \d{\partial}
\def \lam{\lambda}

\begin{document}
\draft
\title{The w-mode instability of ultracompact relativistic stars}

\author{K.D. Kokkotas}
\affiliation{Department of Physics, Aristotle University of
Thessaloniki, Thessaloniki 54124, Greece}

\author{J. Ruoff}
\affiliation{Department of Physics, Aristotle University of
Thessaloniki, Thessaloniki 54124, Greece}

\author{N. Andersson}
\affiliation{Department of Mathematics, University of Southampton,
Southampton SO17 1BJ, UK}

\date{\today}

\begin{abstract}
In this Letter we show that the, mainly spacetime, w-modes may
become unstable in a rotating ultracompact ($R<3M$) relativistic
star. We provide results for the axial modes of a rotating star in
the slow rotation approximation, and demonstrate that an
instability may become active in stars with an ergosphere. These
results suggest that the so-called ``ergoregion instability''
discussed by other authors is, in fact, associated with the
w-modes of the star. The unstable modes are found to grow on a
timescale of seconds to minutes. We discuss briefly whether the
instability is likely to have any astrophysical relevance.
\end{abstract}

\maketitle

{\em Introduction}.---
Just like a rotating black hole, a spinning
neutron star can develop an ergosphere.
The ergosphere is defined as a region where all
observers must  be dragged along with the rotation, and
no observer can remain at rest. Mathematically, the ergosphere is
distinguished as the part of a
stationary spacetime where a spacelike Killing vector becomes
timelike. Hence, the boundary of the ergosphere is defined by the
change in the sign of the $[{tt}]$ component of the metric.
In the case of relativistic stars, an ergosphere may develop inside
in the stellar fluid if the star is sufficiently compact ($R<3M$).

It is by now well known that a neutron star spacetime with an
ergosphere may become unstable. This was first demonstrated by
Friedman~\cite{Friedman78} for scalar and electromagnetic
perturbations. The problem was studied in  detail by Comins and
Schutz \cite{CS78} who considered scalar fields in the background
of a slowly rotating star. Their approach had the advantage that
the problem could be reduced to one dimension, which facilitated
the use of the WKB approximation to calculate the unstable modes
of oscillation (for large $m$, where the perturbations are assumed
to be proportional to $e^{im\phi}$). Comins and Schutz found that
the growth time of the instability would be very large (of the
order of the age of the universe). This would mean that, even if
sufficiently compact and rapidly rotating stars would form, the
instability would not be astrophysically important. More recently,
the problem was revisited by Yoshida and Eriguchi \cite{YE96} who
developed a new numerical scheme which allowed them to extended
the results of Comins and Schutz to smaller values of $m$. Their
results showed that the low-order (in $m$) modes would grow much
faster than indicated by the asymptotic results of Comins and
Schutz.

Since the first studies of the ergoregion instability were carried
out the existence of pulsation modes that are essentially due to
the spacetime itself has been established \cite{KS92}. These are
known as the w-modes \cite{KS92}, and they only rely on the
background curvature generated by the fluid in order to exist. The
actual coupling to the stellar fluid is weak, and the mode mainly
excites oscillations of the spacetime. An extreme case concerns
axial perturbations of a non-rotating perfect fluid star. In this
case no non-trivial fluid modes of oscillations exist. Yet they
have an infinite set of w-modes. A similar set of spacetime modes
exists for polar perturbations, although in that case there are
also non-trivial fluid modes (for example, the f-mode and the
acoustic p-modes), see \cite{AKK96} for an exhaustive discussion.
In general, the w-modes have high frequencies, with the
fundamental mode in the range 5-12kHz,  and very short damping
times, of the order of a tenth of a millisecond. Before the
w-modes were discovered Chandrashekhar and Ferrari \cite{CF91}
suggested that ultracompact stars (with radius smaller than $3M$)
can support oscillation modes due to the trapping of gravitational
waves by the curvature potential (familiar from studies of
perturbed black holes). Detailed studies \cite{Kokkotas94,AKK96}
have shown that these long-lived modes are, in fact, the first few
w-modes of an ultracompact star. The key difference is that,
because these modes are trapped inside the peak of the curvature
potential their damping can be very slow (the damping time is
roughly proportional to the width of the potential barrier).

As we will demonstrate below, these trapped w-modes may
become unstable if the star is set into rotation. We will show that
this instability only becomes active after the star has developed
an ergosphere. This means that the gravitational-wave version of the
ergoregion instability  \cite{Friedman78,CS78,YE96}  sets in through the
spacetime w-modes.

{\em The w-mode instability}.--- In order to establish the
existence of the w-mode instability we study the effect of
rotation on the axial w-modes within the slow rotation formalism.
In writing down the perturbation equations we will keep only terms
of first order in the angular frequency of the star ($\Omega$).
Hence,  we account for the relativistic frame dragging, but do not
consider the rotational deformation of the star (which is an order
$\Omega^2$ effect). In addition, we neglect the coupling between
axial and polar perturbations.

If we focus on purely axial perturbations, the perturbed metric
can be written as
\begin{equation}\label{metric}
  ds^2 = ds^2_0 + 2\sum_{l,m}\(h_0^{lm}(t,r)dt + h_1^{lm}(t,r)dr\)
  \(-\sin^{-1}\theta \d_\phi Y_{lm}d\theta
  + \sin\theta\d_\theta Y_{lm}d\phi\)\;,
\end{equation}
where $Y_{lm} = Y_{lm}(\theta,\phi)$ denote the scalar spherical
harmonics. The unperturbed metric $ds^2_0$ represents a
non-rotating stellar model with a first order rotational
correction $\omega(r)$ (the frame dragging) in the
$[t,\phi]$-component:
\begin{equation}
  ds^2_0 = -e^{2\nu} dt^2 + e^{2\lam} dr^2
  + r^2\(d\theta^2 + \sin^2\theta d\phi^2\)
  - 2\omega r^2\sin^2\theta dtd\phi\;.
\end{equation}
The fluid is assumed to rotate with uniform angular velocity
$\Omega$, and the axial component of the fluid velocity perturbation
can be expanded as
\begin{equation}\label{fluid}
  4\pi(p + \eps)\(\delta u^\theta, \delta u^\phi\)
  = e^{\nu}\sum_{l,m}U^{lm}(t,r)
  \(-\sin^{-1}\theta \d_\phi Y_{lm} \ , \ \sin\theta\d_\theta Y_{lm}\)\;.
\end{equation}
With these definitions Einstein's field equations reduce to three
equations for the three variables $h_0^{lm}$, $h_1^{lm}$ and
$U^{lm}$ \cite{Kojima92}. An equivalent set of equations in the
ADM-formalism is given in \cite{RK02,RSK02}. Assuming a harmonic
time dependence $e^{i\sigma t}$, the perturbation equations can be
written as two ODEs for $h_0$ and $h_1$ and an algebraic relation
for $U$;
\begin{eqnarray}
h_1'&=& \left[\lambda' -\nu'-
{m\omega'(\Lambda-2)\over{\Lambda(\sigma-m\omega)}}\right]h_1
-\left[{2m^2r^2 \over \Lambda}\left( {\omega'^2e^{-2\lambda}\over
{\Lambda(\sigma-m\omega)}} +{{16\pi\varpi^2(p+\eps)}\over
{2m\varpi+\Lambda(\sigma-m\Omega)}}\right)+(\sigma-m\omega)\right]ih_0
\label{h1}
\\
h_0'&=& 2\left[{1\over r}-{m\omega'\over
{\Lambda(\sigma-m\omega)}}\right]h_0
+\left[{{e^{2\nu}(\Lambda-2)}\over
{r^2(\sigma-m\omega)}}-e^{4(\nu-\lambda)}(\sigma-m\omega)\right]ih_1
\label{h0}
\\
U&=&4\pi e^{-2\nu}(p+\eps)\left[ 1- {2 m\varpi \over 2m\varpi +
\Lambda (\sigma-m\Omega)}\right]h_0 \label{U}
\end{eqnarray}
where $p$ is the pressure, $\eps$ the energy density and
$\Lambda=l(l+1)$.

Equations \eqref{h1} and \eqref{U} become singular for certain
values of $\sigma$. This singularity is due to the existence of a
continuous spectrum \cite{Kojima98,BK99,RK02}. The existence of
this continuous spectrum has no impact on our study of the
w-modes, while it greatly affects the study of r-modes.

Given the above equations it is worth noticing that in equation
\eqref{h1} the coefficient of $h_0$ includes terms which are
formally of second order in the rotation rate $\Omega$. Within the
slow rotation approximation these terms can be neglected. However,
this would not be consistent for inertial modes (like the r-mode)
since their frequencies are such that $\sigma\sim \Omega$. Only in
this way can the correct growth times of the r-modes be calculated
\cite{RK02}. For the w-modes we have typically $\sigma \gg \Omega$
and we can neglect second order terms in $\Omega$. However,  we
nevertheless  use the above set of equations in order to be able
to compute both r- and w-modes with the same numerical code.

\begin{figure}[h]
\centering
\includegraphics[height=5cm,clip]{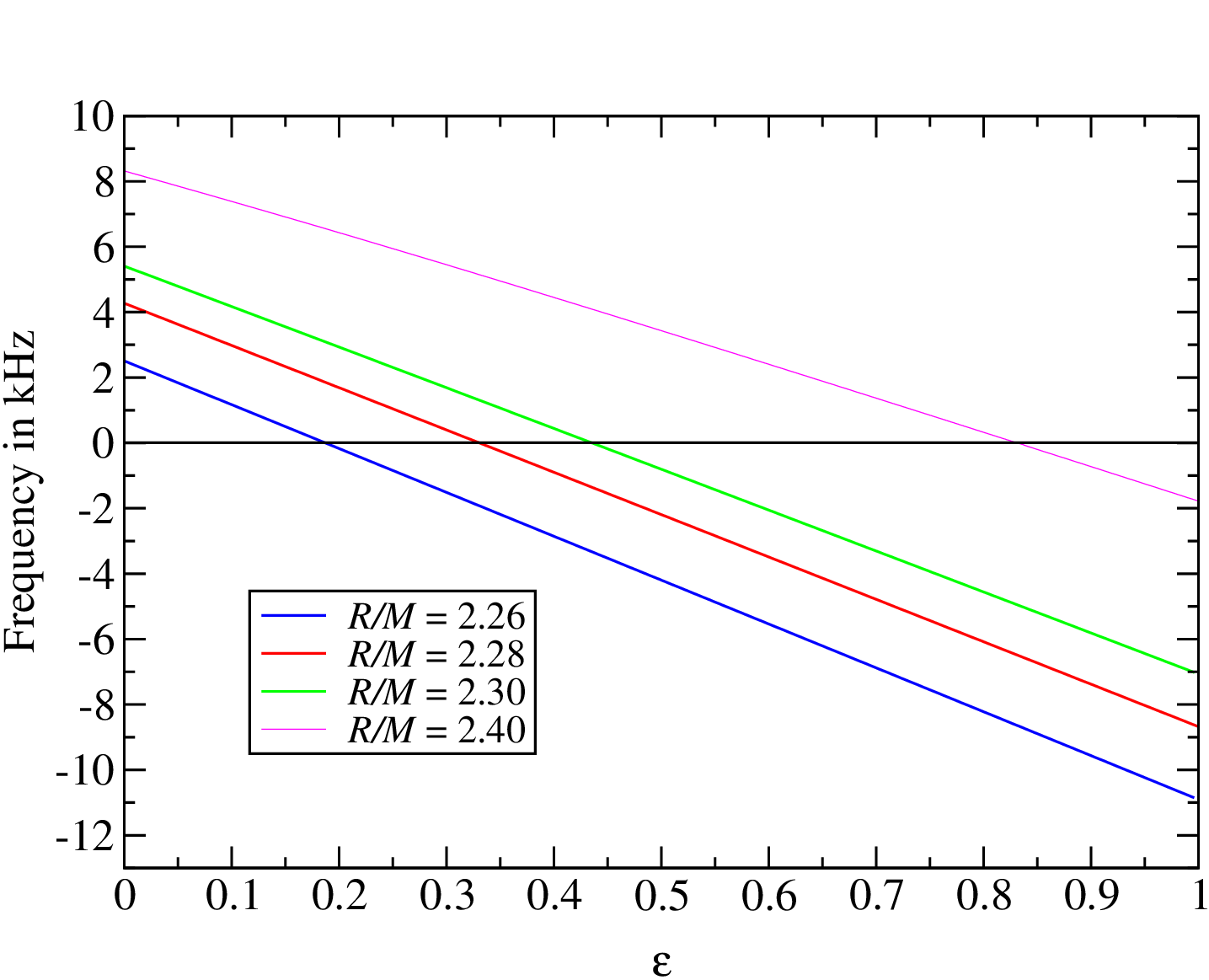}
\includegraphics[height=5cm,clip]{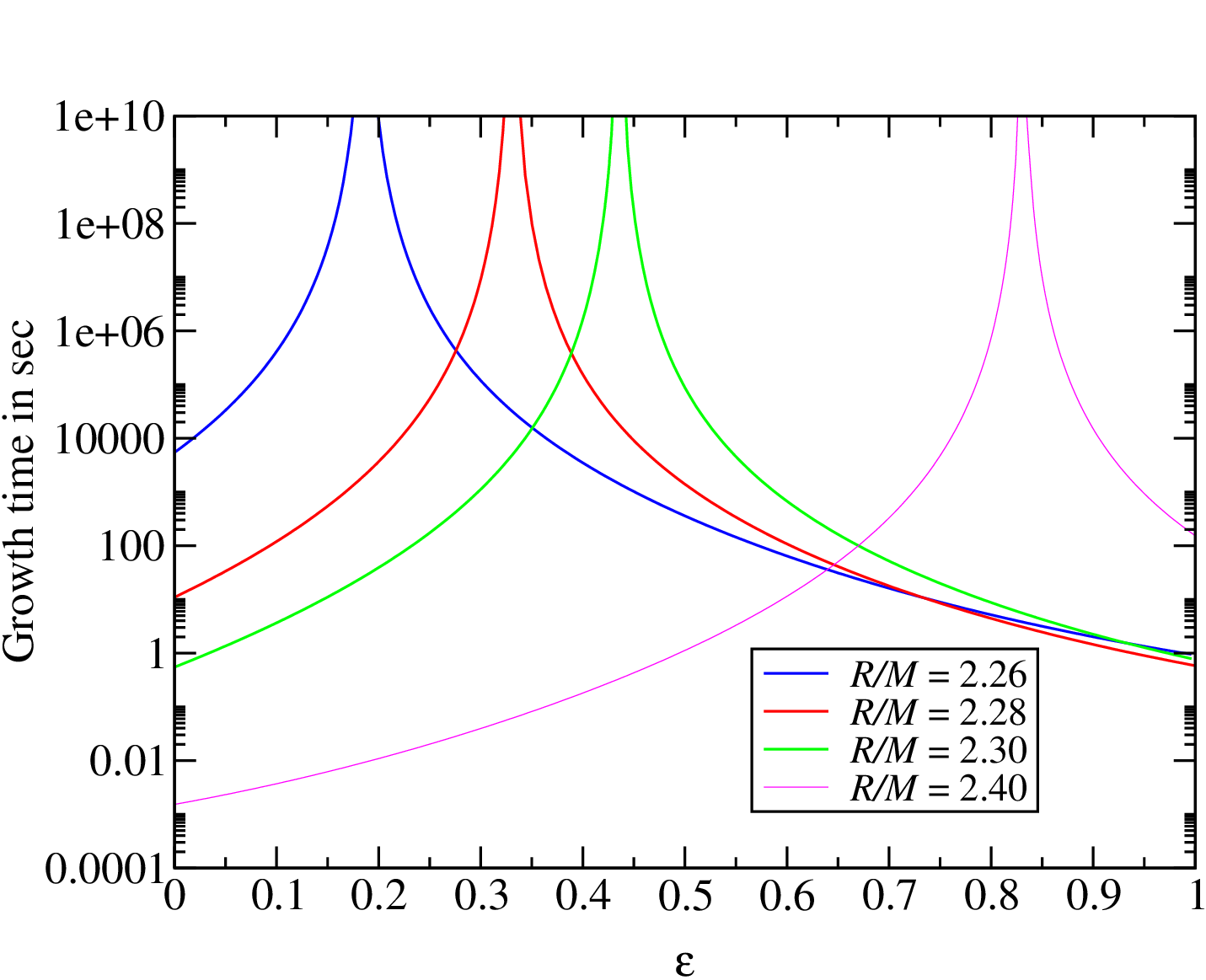}
\caption{Oscillation frequency (left frame) and damping/growth
time (right frame) of the first w-mode of ultracompact stars. The
rotation rate of the star is represented by
$\veps=\Omega/\Omega_K$. The w-mode instability is active beyond
the point where a neutral mode exists, i.e. whenever the mode
frequency is negative. As is clear from the data, the instability
can grow on a timescale of a few tenths of a second to minutes in
very rapidly rotating stars.  } \label{fig1}
\end{figure}

Our numerical results for ultracompact stars are illustrated in
Fig.~\ref{fig1}. The two frames show the oscillation frequency and
damping/growth time of the first w-mode for a star with mass $1.4
M_\odot$, the canonical value for a neutron star and four
different compactness ratios. The illustrated mode is such that
its pattern moves backwards with respect to the rotation as
$\Omega\to 0$. As the rotation rate, represented by $\veps
=\Omega/\Omega_K$ (where $\Omega_K$ is the mass shedding limit),
increases the frequency of the mode decreases and passes through
zero at a critical value. Beyond this point the emerging
gravitational waves drive the mode unstable. The onset of
instability is also signalled by a singularity in the growth time
since a marginally unstable mode would not radiate at all. These
singularities are obvious in the right frame of Fig.~\ref{fig1}.

It is relevant to compare the critical rotation rate at which the
first w-mode becomes unstable to the Kepler limit. For the cases
shown in Fig.~\ref{fig1} we find that the instability becomes
active at $0.19\Omega_K$ for $R/M=2.26$,  at $0.33\Omega_K$ for
$R/M=2.28$, at $0.44\Omega_K$   for $R/M=2.30$ and at
$0.83\Omega_K$ for $R/M=2.40$. In other words, for stars near the
absolute limit of compactness allowed by General Relativity the
w-modes become unstable already at quite low rates of rotation
(below 20\% of the mass-shedding limit). As is clear for the right
frame in Fig.~\ref{fig1} the associated growth times can be as
short as a few tenths of seconds to minutes. Thus the growth time
of the w-mode instability can be very short (Our results indicate
that the unstable gravitational perturbations grow at least four
orders of magnitude faster than the scalar field perturbations
considered by eg Yoshida and Eriguchi~\cite{YE96}. Nevertheless,
we find that the unstable w-modes grow slower than the $l=m=2$
r-mode in all cases we have considered. We have also verified that
the growth times scale with the rotation rate as $\sim
\Omega^{2l+2}$ as one would expect.

The class of ``trapped'' w-modes which becomes unstable consists
of a finite number of modes. The number of the ``trapped'' modes
increases with the compactness of the star and several of these
modes become unstable for higher rotational rates. For example for
a star with compactness $R/M=2.26$ near the mass shedding limit
more than 4 modes become unstable.

\begin{figure}[h]
\centering
\includegraphics[height=5cm,clip]{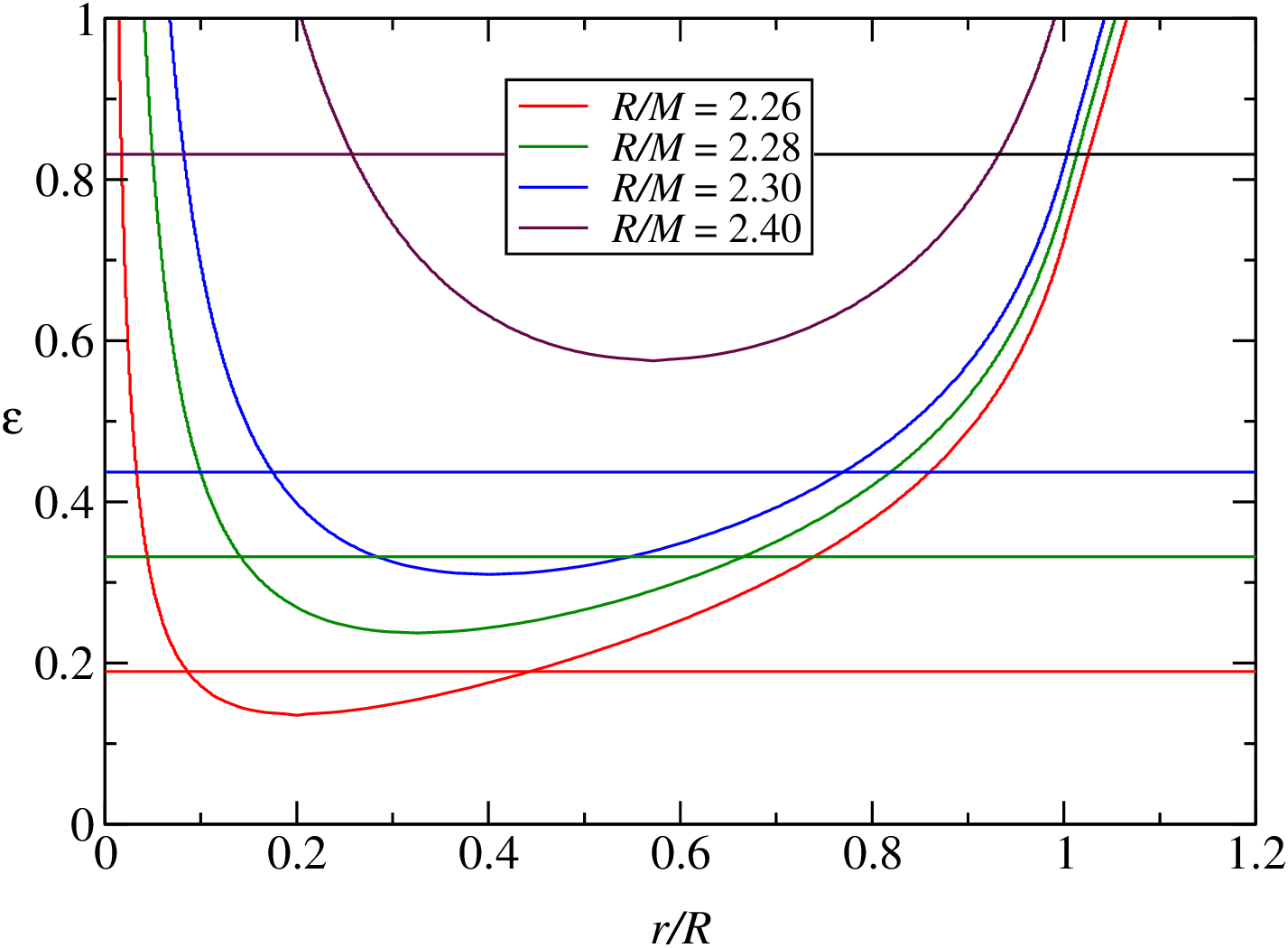}
\caption{The size of the ergosphere for our four stellar models is
shown for different values of $\veps$. The ergosphere is present
above a certain critical value of $\veps$. We also show (as
straight lines) the critical value of $\veps$ at which the first
w-mode becomes unstable for each stellar model. As is clear from
this data, the w-mode instability only becomes active after an
ergosphere has formed.} \label{fig2}
\end{figure}

We now want to establish that the w-modes will only be unstable in stars
that have developed an ergosphere.
To test the existence of an ergosphere we follow Schutz and
Comins~\cite{SC78} and approximate $g_{tt}$ as
\begin{equation}
  \label{eq:gtt}
  g_{tt} = -e^{2\nu} + \omega^2r^2\sin^2\theta
\end{equation}
Given this,
an ergosphere is present whenever $g_{tt} > 0$, i.e.  restricting
ourselves to the equatorial plane $\theta = \pi/2$ we require
\begin{equation}
  \label{eq:ergo}
  e^{\nu} < r\omega\;.
\end{equation}
Figure~\ref{fig2} illustrates the  ergosphere (in the equatorial
plane) for our ultracompact stellar models and varying rates of
rotation. Also shown in the figure are the critical values of the
rotation parameter $\veps$ at which the first w-mode becomes
unstable. As is clear from the figure, the w-mode only becomes
unstable once the star has developed an ergosphere.

{\em Discussion}--- In this Letter we have demonstrated that the
spacetime w-modes may become unstable in ultracompact rotating
stars. Our results show that the instability come into operation
after the star has developed an ergosphere, and that the unstable
modes can grow quite fast --- on a timescale of seconds to
minutes. The w-mode instability is a variation of the
gravitational-wave driven (CFS) instability, which was discovered
by Chandrasekhar \cite{Chandra70} for the Maclaurin spheroids and
subsequently investigated in detail by  Friedman and
Schutz~\cite{FS78}. The key difference between the w-mode
instability and the standard CFS instability of fluid oscillation
modes (like the axial r-mode \cite{AK01}) is that the w-modes are
mainly due to spacetime oscillations. They are gravitational waves
trapped (in the present case) inside the peak of the curvature
potential. Their instability can be understood from the fact that
an oscillation mode can be associated with a negative energy
provided that the star has an ergosphere.

As an interesting aside it is worth commenting on the fact that
despite them being endowed with an ergosphere, rotating black holes
do not suffer an analogous instability. The reason for this is simple:
In the black hole case, the leakage of energy through the horizon
proceeds fast enough to stabilize the system.

The demonstration that the w-modes may become unstable is
interesting from a conceptual point of view. However, the
astrophysical relevance of this instability is questionable. The
main reason for this is that it is very difficult to construct
realistic stable stellar models which would exhibit an
ergosphere~\cite{BI76,KEH89}. Indeed, there are very few
``realistic'' supranuclear equations of state that permit stable
stellar models significantly more compact than $R\approx 3M$. A
possibility that cannot yet be completely ruled out is that of
ultracompact strange stars.  One can, in principle, generate very
compact quark stars but we have not yet any observational evidence
for the existence of such objects.

Finally, it is worth discussing two features that make the w-mode
instability differ somewhat from the more familiar CFS instability
of fluid oscillations modes. The first concerns dissipation. Since
the w-modes are spacetime modes that hardly excite any fluid
motion {\em there are no apparent dissipation mechanisms}. The
viscous damping of the instability will certainly be significantly
weaker than that of (say) the r-modes since the viscosity will
only act on a relatively small part of the w-mode energy. The
second point concerns the saturation of the unstable mode. Recent
work suggests that the unstable r-modes saturate at a relatively
low amplitude because of strong coupling to short wavelength
inertial modes. The weak coupling to the stellar fluid in the
present case may mean that this mechanism will be significantly
less efficient, if of any relevance, in the case of unstable
w-modes. In fact, one could argue that these modes ought to
saturate because of nonlinear spacetime effects. More detailed
investigations into these issues would be of interest since the
may help improve our understanding of nonlinear dynamical
spacetimes.

{\em Acknowledgements}.---  J.R.~is supported by the Marie Curie
Fellowship No.~HPMF-CT-1999-00364. This work has been supported by
the EU Programme 'Improving the Human Research Potential and the
Socio-Economic Knowledge Base' (Research Training Network Contract
HPRN-CT-2000-00137). NA is a Philip Leverhulme prize fellow.

\end{document}